\documentclass[12pt]{article}
\usepackage{amssymb,latexsym}
\usepackage[intlimits,sumlimits]{amsmath}
\usepackage{hyperref}
\usepackage{amsthm}
\usepackage{accents}
\usepackage{cite}
\usepackage{cleveref}
\usepackage{xfrac}
\usepackage{graphicx}
\usepackage{caption}
\usepackage{color}
\usepackage[bottom]{footmisc}

\usepackage[normalem]{ulem}
\numberwithin{equation}{section}

\newcommand{\R}{\mathbb{R}}
\renewcommand{\o}{\mathrm{odd}}
\newcommand{\e}{\mathrm{even}}

\newcommand{\definedas}{\mathrel{\raise.095ex\hbox{\rm :}\mkern-5.2mu=}}
\newcommand{\asdefined}{\mathrel{=\mkern-5.2mu}\raise.095ex\hbox{\rm :}\;}

\newtheorem{Thm}{Theorem}[section]

\newtheorem{Def}[Thm]{Def}

\begin{document}

\title{Coordinates are messy\\ --- not only in General Relativity}
\author{Carla Cederbaum\thanks{cederbaum@math.uni-tuebingen.de} \\Mathematics Department \\Eberhard Karls Universit\"at T\"ubingen
\\ \\ Melanie Graf\thanks{melanie.graf@uni-hamburg.de} \\ Mathematics Department \\Eberhard Karls Universit\"at T\"ubingen/\\ now Universit\"at Hamburg}
\date{}

\maketitle

\abstract{The coordinate freedom of General Relativity makes it challenging to find mathematically rigorous and physically sound definitions for physical quantities such as the center of mass of an isolated gravitating system. We will argue that a similar phenomenon occurs in Newtonian Gravity once one ahistorically drops the restriction that one should only work in Cartesian coordinates when studying Newtonian Gravity. This will also shed light on the nature of the challenge of defining the
center of mass in General Relativity. Relatedly, we will give explicit examples of asymptotically Euclidean relativistic initial data sets which do not satisfy the Regge--Teitelboim parity conditions often used to achieve a satisfactory definition of center of mass. These originate in our joint work~\cite{WIP} with Jan Metzger. This will require appealing to Bartnik's asymptotic harmonic coordinates.}

\section{Preferred Systems of Coordinates (or not)}\label{sec:intro}
As we all know, Euclidean space --- the stage of Newtonian Gravity
--- knows preferred systems of coordinates, called Cartesian
coordinates. In such coordinates, the Euclidean metric $\delta$ takes
its canonical form. Similarly, the Minkowski spacetime --- the setting
of Special Relativity --- carries preferred systems of coordinates in
which the Minkowski metric $\eta$ takes its canonical form. In
contrast, curved spacetimes --- the mathematical framework of General
Relativity --- and initial data sets therein are well-known not to
admit any 'canonical' or 'preferred' coordinates in general. This freedom in the choice of coordinates makes it challenging to find
mathematically rigorous and physically sound definitions for physical
quantities such as the center of mass of an isolated gravitating
system in General Relativity as is well-known and will be discussed in this article. We will
argue that a similar phenomenon occurs in Newtonian Gravity once one
ahistorically and somewhat unnecessarily drops the restriction that
one should only work in Cartesian coordinates when studying Newtonian Gravity. This
will also shed light on the nature of the challenge of defining the
center of mass in General Relativity. Relatedly, we will give explicit examples of asymptotically Euclidean relativistic initial data sets which do not satisfy the ``Regge--Teitelboim (parity) conditions'' often used to achieve a satisfactory definition of center of mass. These originate in our joint work~\cite{WIP} with Jan Metzger.

\section{Isolated Systems at a Given Instant of Time}
\label{sec:iso}
Let's begin by recalling the standard definitions of an ``isolated
system at a given instant of time'' in both Newtonian Gravity and General Relativity. We will also
recall the standard definitions of (total) mass of such systems along
the way and discuss the convergence of the involved integrals.

\subsection{Isolated Systems at a Given Instant of Time in Newtonian Gravity}
In Newtonian Gravity, we can think of an ``isolated system at a given instant of time''
as given by a matter density function
$\rho\colon \mathbb{R}^3\to [0,\infty)$ which has compact support or
at least decays suitably fast towards infinity. For example, one could
ask that $\rho=O(r^{-3-\varepsilon})$ as $r\to\infty$ for some
(small) $\varepsilon>0$, that is, $\rho$ decays to zero at least as
fast as $r^{-3+\varepsilon}$, where $r$ denotes the radial coordinate
on $\mathbb{R}^3$. Alternatively but not equivalently, one could ask
that $\rho\in L^1(\mathbb{R}^3)$. Both assumptions are independent of
the chosen Cartesian coordinates because any two systems of Cartesian
coordinates on $\mathbb{R}^3$ differ only by a rigid
motion. 
Either of these decay assumptions is sufficient for the total mass
\begin{align}\label{def:mass}
m=\iiint_{\mathbb{R}^3} \rho(\vec{x})\, d\vec{x}
\end{align}
to be well-defined and finite. Anticipating the discussion
below, let us point out that the
$O$-assumption suggests computing the integral in
\eqref{def:mass} as an improper Riemann integral in polar coordinates,
while the $L^{1}$-assumption suggests treating it as a Lebesgue integral. Of
course, the resulting mass
$m$ will be the same whatever notion of integral one refers to, as
long as it converges. However, taking the former viewpoint, we can
take advantage of cancellations in the spherical
integrals. Also note that the
decay assumptions are of course \emph{not} independent of arbitrary
coordinate changes.

\subsection{Isolated Systems at a Given Instant of Time in General Relativity}
In General Relativity, an ``isolated system at a given instant of time'' is modelled as
an asymptotically Euclidean (or asymptotically flat) relativistic initial data set (or time-slice):
As usual, an initial data set $(M,g,K)$ consists of a
$3$-dimensional Riemannian manifold
$(M,g)$ carrying a symmetric $(0,2)$-tensor field
$K$ playing the role of second fundamental form (or extrinsic
curvature) of the initial data set  in the spacetime
modelling the system.

\begin{figure}[h]
 \begin{center}
\includegraphics[scale=0.7]{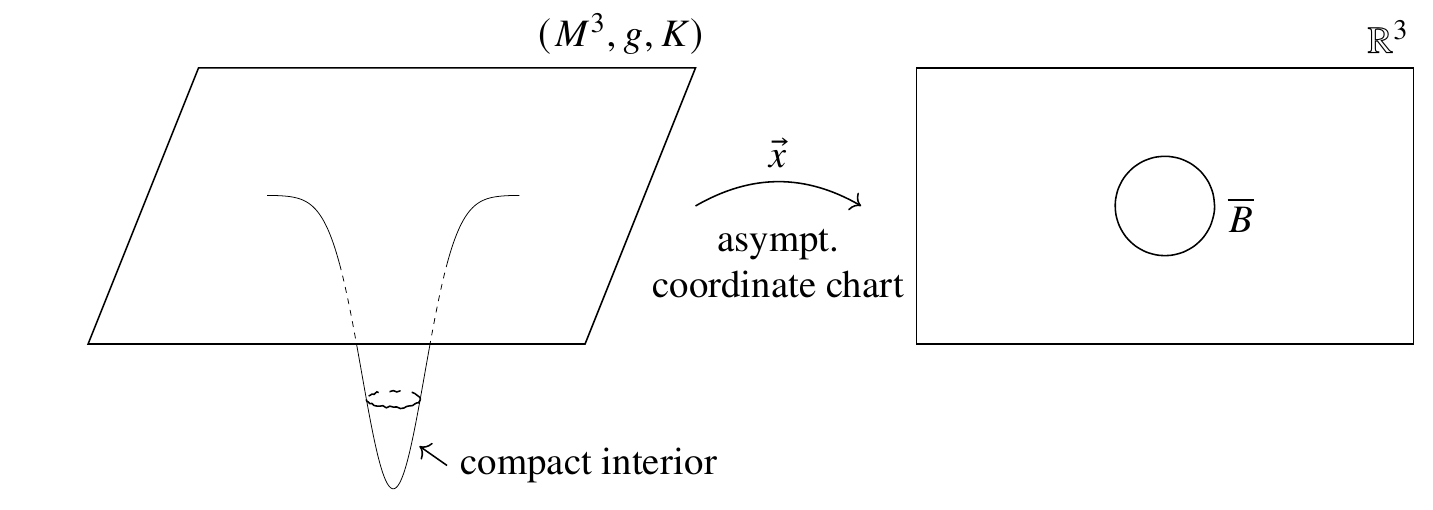}
\end{center}
\caption{An asymptotically Euclidean relativistic initial data set $(M,g,K)$ and the image of its asymptotic end $M\setminus C$ in $\R^{3}$ under the asymptotic coordinate chart $\vec{x}$.}\label{fig:graphic-coord-CoM}
\end{figure}

In addition, a relativistic initial data set carries an
energy density $\mu\colon
M\to\mathbb{R}$ and a momentum density one-form $J$ related to
$g$ and
$K$ via the well-known Einstein constraint equations 
\begin{align}\label{eq:Hconstraint}
 \operatorname{R} - \vert K\vert ^2 + (\operatorname{tr} K)^2 &=2\mu\\\label{eq:Mconstraint}
\operatorname{div} (K - (\operatorname{tr} K) g)&=-J.
\end{align}
and derived from the energy-momentum tensor $T$ of the spacetime. Here, $\operatorname{R}$ denotes the scalar curvature of $(M,g)$, and $\vert\cdot\vert$, $\operatorname{tr}$, and $\operatorname{div}$ denote the tensor norm, trace, and divergence  with respect to $g$, respectively. We will adopt the
following standard definition, see also \Cref{fig:graphic-coord-CoM}.
\begin{Def}[Asymptotically Euclidean Relativistic Initial Data Set]\label{def:AE} A relativistic initial data set $(M,g,K)$ with energy density $\mu$ and momentum density $J$ is called \emph{asymptotically Euclidean} if there exists a compact set $C\subset M$, a radius $R>0$, and an \emph{asymptotic coordinate chart $\vec{x}\colon M\setminus C \to \R^3\setminus B_R(0)$} such that  
\begin{align}\label{eq:gdecay}
g_{ij}-\delta_{ij} &= O_2(r^{-\frac{1}{2}-\varepsilon})\\\label{eq:Kdecay}
K_{ij} &= O_1(r^{-\frac{3}{2}-\varepsilon})\\\label{eq:muJdecay}
\mu, J_i &= O_0(r^{-3-\varepsilon})
\end{align}
as $r=\vert\vec{x}\,\vert\to\infty$ for some decay parameter $\varepsilon>0$, where $g_{ij}$, $K_{ij}$, and $J_i$ denote the components of $g$, $K$, and $J$ in the coordinates $\vec{x}$, respectively. Alternatively but not equivalently, one can replace Assumption \eqref{eq:muJdecay} by asking that $\mu, J_i\in L^1(\R^3\setminus B_R(0))$.

Here, the index $k\in\mathbb{N}_0\cup\{\infty\}$ in $O_k(r^\alpha)$ for some $\alpha<0$ is a shorthand for asking that derivatives of order up to $k$ decay 'accordingly' as $r\to\infty$, that is, first derivatives decay to zero at least as fast as $r^{\alpha-1}$, second derivatives decay to zero at least as fast as $r^{\alpha-2}$, etc. In what follows, we will slightly abuse notation and extend this to the 'decay rate' $\alpha=0$, so that $f=O_k(r^0)$ will mean that the function $f$ stays bounded as $r\to\infty$, while its order $l$ derivatives decay to zero at least as fast as $r^{-l}$ as $r\to\infty$ whenever $l\leq k$.
\end{Def}

The (total) mass $m_{\text{ADM}}$ of an asymptotically Euclidean relativistic initial data set $(M,g,K)$ was defined by Arnowitt, Deser, and Misner in \cite{ADM1962} via the (total) energy $E_{\text{ADM}}$ and (total) linear momentum $\vec{P}_{\text{ADM}}$ and has become the standard definition, satisfying many desirable properties such as for example positivity (\cite{SchoenYau,Witten}):
 \begin{align}\label{def:E}
 E_{\text{ADM}}&\definedas\frac{1}{16\pi } \lim_{R\to \infty} \iint_{S_R(0)} \sum_{i,j=1}^{3} (\partial_ig_{ij}-\partial_jg_{ii})\frac{x^j}{R} dA_\delta,\\\label{def:P}
 P_{\text{ADM}}^j &\definedas \frac{1}{8\pi } \lim_{R\to \infty} \iint_{S_R(0)} \sum_{i=1}^{3} \left( (\operatorname{tr} K) g_{ij} -K_{ij}\right) \frac{x^i}{R} dA_\delta,\\\label{def:m}
 m_{\text{ADM}}&\definedas\sqrt{E_{\text{ADM}}^2-\left\vert\vec{P}_{\text{ADM}}\right\vert^2},
\end{align}
 where $dA_{\delta}$ denotes the Euclidean area measure on $S_R(0)$. In \cite{Bartnik86}, Bartnik uses harmonic asymptotically Euclidean coordinate charts to prove that $E_{\text{ADM}}$ is well-defined and independent of the choice of asymptotically Euclidean coordinate charts, see \Cref{sec:comparison}. Note that Bartnik uses the weaker $L^{1}$-decay condition on $\mu$. From this, \eqref{def:P}, \eqref{eq:Kdecay} and \eqref{eq:muJdecay}, it follows directly that $m_{\text{ADM}}$ is also well-defined and independent of the choice of asymptotically Euclidean coordinate chart via the Divergence Theorem. Again, this argument only uses the weaker $L^{1}$-condition on~$J$.
 
You may wonder why one asks for such general decay rates in \eqref{eq:gdecay}--\eqref{eq:muJdecay}, and not just for, say,
\begin{align}
g_{ij}-\delta_{ij} = O_2(r^{-1})
\end{align}
etc.. This has two reasons: First, the theory of asymptotically Euclidean relativistic initial data sets does not become more complicated if one does so. Second, it actually becomes richer, i.e., allows for more examples, see \cite{WIP} and the references cited therein. 

Next, let us discuss how a single relativistic initial data set can carry different asymptotic coordinate charts and discuss the relationship between different such charts.

 \section{Comparing Different Asymptotic Coordinate Systems}\label{sec:comparison}
Clearly, if a relativistic initial data set is asymptotically Euclidean for some asymptotic coordinate chart $\vec{x}$, it will also be asymptotically Euclidean for any asymptotic coordinate chart $\vec{y}$ arising from $\vec{x}$ by a rigid motion\footnote{\hsize=0.94\textwidth
The careful reader may note that one may need to change the compact set $C$ and the radius $R$ from \Cref{def:AE} for the coordinate chart $\vec{y}$; we will ignore such subtleties in this article for the sake of readability.\label{fn:compact}}, for the same decay parameter~$\varepsilon$. Moreover, the class of possible transformations between two asymptotic coordinate charts $\vec{x}$ and $\vec{y}$ with respect to which a given relativistic initial data set $(M,g,K)$ is asymptotically Euclidean is much richer than just rigid motions, even when fixing the decay parameter $\varepsilon$. Here is an example: Let $(M,g,K)$ be a relativistic initial data set with energy and momentum densities $\mu$ and $J$ which is asymptotically Euclidean with respect to some asymptotic coordinate chart $\vec{x}$ and for some decay parameter $\varepsilon>0$. Then $(M,g,K)$ will also be asymptotically Euclidean with decay parameter $\varepsilon$ with respect to the asymptotic coordinate chart
\begin{align}\label{eq:coordchange}
\vec{y}\definedas \vec{x}+\sin(\ln r)\,\vec{a}
\end{align}
for some non-vanishing $\vec{a}\in\R^3$, with $r=\vert\vec{x}\vert$ as
before. This can be seen by a straightforward computation. In
particular, note that 
\Cref{eq:coordchange} is very similar to a mere translation, differing
only in the bounded factor $\sin(\ln r)=O_{\infty}(r^{0})$ as $r\to\infty$.

On the other hand, it crucially depends on the choice of asymptotic coordinate chart whether a given relativistic initial data set ``is'' asymptotically Euclidean: For example, any relativistic initial data set which is asymptotically Euclidean with respect to an asymptotic coordinate chart $\vec{x}$ will not be asymptotically Euclidean with respect to the chart $\vec{y}\definedas 2\vec{x}$ as one easily computes. We will hence refer to an asymptotic coordinate chart $\vec{x}$ as an \emph{asymptotically Euclidean coordinate chart} for a given relativistic initial data set $(M,g,K)$ (with energy and momentum densities $\mu$ and $J$) if $(M,g,K)$ is asymptotically Euclidean with respect to $\vec{x}$.\newpage
 
Summarizing, the class of asymptotically Euclidean coordinate charts for a given relativistic initial data set is much richer than the class of Cartesian coordinate systems on Euclidean space. This applies in particular to the Euclidean relativistic initial data set $(M=\R^{3},g=\delta,K=0)$ sitting inside the Minkowski spacetime. Here, one sees that the Cartesian coordinate systems are asymptotically Euclidean coordinate charts, but by far not the only asymptotically Euclidean coordinate charts. 

\subsection{Divergence of Mass}\label{sec:DS}
 At this point, it is instructive to recall that the decay condition \eqref{eq:gdecay} cannot be relaxed as was shown by a counter-example by Denissov and Solovyev \cite{DenSol1983}: Inspired by their example, let us consider the Euclidean relativistic initial data set $(M=\R^{3},g=\delta,K=0)$ in the coordinates 
 \begin{align}
 \vec{y}\definedas \left(1+\frac{a}{\sqrt{r}}\right)\, \vec{x}
 \end{align}
 for some non-zero $a\in\R$ which leads to \eqref{eq:gdecay} with $\varepsilon=0$ and the unphysical result $m_{\text{ADM}}=\frac{a^{2}}{8}$. One can argue similarly for \eqref{eq:Kdecay} as we will discuss elsewhere; alternatively, one can compute in a lengthy but straightforward way that the decay conditions \eqref{eq:gdecay} and \eqref{eq:Kdecay} transform equivariantly under coordinate boosts in the ambient spacetime. From this and the example by Denissov and Solovyev, one can conclude that \eqref{eq:Kdecay} is necessary for physicality of the definition of $m_{\text{ADM}}$.

In summary, (convergence and coordinate independence of) mass is very well understood in both Newtonian Gravity and General Relativity and depends crucially on the decay of the matter variables, as well as, in General Relativity, on the asymptotics of the relativistic initial data set itself. 

One of the main tools introduced by Bartnik for the study of mass and energy are the ``harmonic asymptotically Euclidean coordinates'' we will now explain.

\section{A Canonical Choice: Harmonic Coordinates}\label{sec:harmonic}
Cartesian coordinates are not only canonical for the Euclidean metric, they are also \emph{harmonic}, that is, they satisfy the system of partial differential equations
\begin{align}
\triangle_{\delta} \vec{x}=0,
\end{align}
a shorthand for the system of equations
\begin{align}
\triangle_{\delta} x^{i}=0 \text{ for }i=1,2,3,
\end{align}
where $\triangle_{\delta}$ denotes the Euclidean Laplacian. 

Exploiting this insight, Bartnik showed in \cite{Bartnik86} that  asymptotically Euclidean relativistic initial data sets $(M,g,K)$ always possess \emph{harmonic} asymptotically Euclidean coordinate charts, that is, asymptotically Euclidean coordinate charts satisfying the geometric system of partial differential equations
\begin{align}
\triangle \vec{x}=0,
\end{align}
where $\triangle$ denotes the Laplacian with respect to $g$. Here, ``geometric'' means that the partial differential equations themselves do not depend on a choice of (local or asymptotic) coordinate chart.

Furthermore, Bartnik showed \cite[Theorem 3.1]{Bartnik86} that two such harmonic asymptotically Euclidean coordinate charts $\vec{x}$, $\vec{y}$ are related by a rigid motion up to suitably lower order terms, 
\begin{align}
\vec{y}=Q\vec{x}+\vec{a}+O_{0}(r^{\frac{1}{2}-\varepsilon}),
\end{align}
for a special orthogonal matrix $Q\in\R^{3\times3}$ and a vector $\vec{a}\in\R^{3}$. In particular, there are \emph{more} harmonic asymptotically Euclidean coordinate charts on Euclidean space than just the Cartesian coordinate systems: for example, the coordinates $\vec{y}\definedas \vec{x}+\frac{\vec{b}}{r}$ for some non-trivial vector $\vec{b}\in\R^{3}$ are also harmonic. 

\section{On the Center of Mass of Isolated Systems at a Given Instant of Time}
Let us now move on to the definition of (total) center of mass, where the situation is somewhat drastically different than for energy, linear momentum, and mass.
Again, we will first take a look at the (total) center of mass of an isolated system at a given instant of time in Newtonian Gravity.

\subsection{On the Center of Mass in Newtonian Gravity}
 Provided $m\neq 0$, the center of mass in Newtonian Gravity is naturally defined as the averaged weighted integral of the position vector $\vec{x}$,
\begin{align}\label{eq:CNG}
\vec{C}\definedas \frac{1}{m} \iiint_{\mathbb{R}^3} \rho(\vec{x}) \,\vec{x}\, d\vec{x}.
\end{align}
Looking at \eqref{eq:CNG} as a Lebesgue integral, it suggests itself that one should ask that $\rho r \in L^{1}(\R^{3})$. Instead, for this to be well-defined and finite as an improper Riemann integral,
\begin{align}\label{eq:improper}
\vec{C}=\lim_{R\to \infty} \int_{0}^R \iint_{S_r(0)} \rho(\vec{x})\, \vec{x}\,dA_{\delta}\,dr,
\end{align}
it suggests itself to assume $\rho=O_{0}(r^{-4-\varepsilon})$ for some $\varepsilon>0$, in analogy with the choice of decay rate used for ensuring that the mass is well-defined; this of course settles the convergence issue for Newtonian gravitating systems. 

However, let us --- ahistorically --- take a different approach in analogy with the standard approach taken to resolve the corresponding issue in General Relativity. To this end, let us instead make a further parity-based decay assumption, namely 
\begin{align}
\rho^\o=O_0(r^{-4-\varepsilon}),
\end{align}
 where 
\begin{align}
\rho^\o(\vec{x})\definedas \frac{1}{2} (\rho(\vec{x})-\rho(-\vec{x}))
\end{align}
 is the odd part of $\rho$. This approach relies on the insight that the contribution to \eqref{eq:improper} of the even part 
 \begin{align}
 \rho^{\e}\definedas \rho-\rho^{\o}
 \end{align}
  vanishes by parity on each sphere $S_{r}(0)$.\newpage
  
   In view of the analogous approach taken in General Relativity, let us take the time to consider the \emph{parity condition} 
  \begin{align}
  \rho^{\o}=O_0(r^{-4-\varepsilon})
  \end{align}
   and its properties in more detail. Importantly, we would like to bring to the reader's attention that the parity condition is \emph{not} independent of the choice of Cartesian coordinate systems because the reflection $\vec{x}\mapsto-\vec{x}$ involved in the definition of $\rho^{\o}$ does not interact well with translations. However, the desirable invariance under choice of Cartesian coordinate systems can be restored if one assumes that in some, and hence all, Cartesian coordinate systems, one has $\rho =O_1(r^{-3-\varepsilon})$, by appealing to the Mean Value Theorem. 
 
 \subsubsection{Transformation Behavior of the Center of Mass in Newtonian Gravity}\label{sec:transfo}
Of course, when well-defined by asking that $\rho\in L^{1}(\R^{3})$ or $\rho=O_{0}(r^{-4-\varepsilon})$ for some $\varepsilon>0$, the center of mass $\vec{C}$ transforms as expected under changes of Cartesian coordinate systems, which can suggestively be written as 
\begin{align}\label{eq:transfo}
\vec{C}_{\vec{y}}=Q\vec{C}_{\vec{x}}+\vec{a}.
\end{align}
 But what happens if one --- ahistorically --- allows asymptotically Euclidean coordinate charts on the Euclidean stage of Newtonian Gravity? It will be instructive to study this in an explicit example similar to \eqref{eq:coordchange}, i.e., $\vec{y}\definedas \vec{x}+\sin(\ln r)\,\vec{a}$, but modified to obtain a global coordinate chart $\vec{z}$ on $\R^{3}$, 
 \begin{align}\label{eq:z}
 \vec{z}\definedas \vec{x}+\sigma(r)\sin(\ln r)\,\vec{a},
 \end{align}
where $\sigma\colon[0,\infty)\to\R$ is a strictly increasing cut-off function satisfying $\sigma(r)=0$ for $r<2$ and $\sigma(r)=1$ for $r>3$. Computing the center of mass $\vec{C}_{\vec{z}}$ according to \eqref{eq:CNG} with respect to the asymptotically Euclidean coordinate chart $\vec{z}$, for a point particle matter density $\rho(\vec{x})=m\delta(\vec{x})$, one finds $\vec{C}_{\vec{x}}=\vec{0}$ but $\vec{C}_{\vec{z}}$ diverges like $ \sin(\ln s)\vec{a}$ for $s=\vert\vec{z}\vert\to\infty$. 

This can be made mathematically more precise by using a surface integral approach via the Divergence Theorem and the Poisson equation for the Newtonian potential as elaborated by Cederbaum and Nerz~\cite{CederbaumNerz2015}. 

Briefly put, once one ahistorically allows more general asymptotically Euclidean coordinate charts in Newtonian Gravity, the center of mass is \emph{not} generically a well-defined quantity even if $\rho=O_{0}(r^{-4-\varepsilon})$ as $r\to\infty$. From the perspective of Newtonian Gravity arising as the Newtonian limit of General Relativity for slow speeds and small masses, it thus becomes reasonable to expect a similar phenomenon to occur in General Relativity. We will now turn to this.

\subsection{On the Center of Mass in General Relativity}
In General Relativity, a (total) notion of center of mass $\vec{C}_{\text{B\'{O}RT}}$ of an isolated system at a given instant of time was put forward by Beig and \'{O} Murchadha in \cite{BOM87}, based on previous work by Regge and Teitelboim~\cite{ReggeTeitelboim74} and similar in spirit and derivation to the ADM-quantities. For an asymptotically Euclidean relativistic initial data set $(M,g,K)$ with $E_{\text{ADM}}\neq0$, its components are \emph{formally} defined by
\begin{align}\label{eq:CBORT}
C_{\text{BOM}}^\ell\definedas & \frac{1}{16\pi E_{ADM}} \lim_{R\to \infty} \left\{\iint_{S_R(0)} x^\ell \sum_{i,j=1}^{3} (\partial_ig_{ij}-\partial_jg_{ii})\frac{x^j}{R} \right. dA_\delta-\left.\iint_{S_R(0)}\sum_{i=1}^{3} \big(g_{i\ell}\frac{x^i}{R}- g_{ii}\frac{x^\ell}{R} \big)  dA_\delta\right\}
\end{align}
with respect to the given asymptotically Euclidean coordinate chart $\vec{x}$. Just as in the Newtonian case, this is a formal definition in the sense that it need not and does not always converge. \newpage

One instance where it diverges\footnote{\hsize=0.94\textwidth For further examples of divergence of the center of mass, see \cite{CederbaumNerz2015} and the references cited therein.} is the canonical Schwarzschild relativistic initial data set $(\R^{3}\setminus \overline{B_{2m}(0)},\frac{1}{1-\frac{2m}{r}}dr^{2}+r^{2}d\Omega^{2},K=0)$ of mass $m\neq0$, when considered with respect to the asymptotic coordinate chart $\vec{y}$ arising from the Cartesian coordinates $\vec{x}$ computed from the spherical polar Schwarzschild coordinates via~\eqref{eq:coordchange}. As in the Newtonian Gravity case discussed above, one finds via a lengthy computation that $\left(\vec{C}_{\text{B\'{O}RT}}\right)_{\vec{y}}$ diverges like $\sin(\ln s)\vec{a}$ for $s=\vert\vec{y}\vert\to\infty$, while of course $\left(\vec{C}_{\text{B\'{O}RT}}\right)_{\vec{x}}$=0 and $E_{\text{ADM}}=m$. We would like to draw the reader's attention to the fact that this initial data set has $\mu=0$, $J=0$, so the divergence problem clearly does not arise from poor decay of the matter.

A first idea one might have to remedy the divergence problem of $\vec{C}_{\text{B\'{O}RT}}$ could be to assume the stronger decay condition $g_{ij}-\delta_{ij}= O_1(r^{-2-\varepsilon})$, thereby enforcing convergence in a way similar to remedying the convergence issue of $E_{\text{ADM}}$ discovered by Denissov and Solovyev, see \Cref{sec:DS}. However, this implies $E_{\text{ADM}}=0$ which is undesirable when interested in the center of mass. 

Instead, one usually resorts to parity assumptions. Before we do so in \Cref{sec:parity}, let us briefly take a look at the transformation behavior of the center of mass under changes of asymptotic coordinates.

\subsubsection{Transformation Behavior of the Center of Mass in General Relativity}
As in the Newtonian case discussed in \Cref{sec:transfo}, when $\vec{C}_{\text{B\'{O}RT}}$ is well-defined (see below), the center of mass $\vec{C}_{\text{B\'{O}RT}}$ transforms as expected under ``asymptotic Euclidean motions'', i.e., under changes of asymptotic coordinate systems that can be written as
\begin{align}
\vec{y}=Q\vec{x}+\vec{a}
\end{align}
with $Q$ and $\vec{a}$ as before. That is to say that \eqref{eq:transfo} holds also in the relativistic case.

\subsubsection{Introducing the Regge--Teitelboim Parity Conditions}\label{sec:parity}
As hinted to in the Newtonian Gravity discussion above, the standard way out of the divergence dilemma is to assume parity conditions as suggested by Regge and Teitelboim in \cite{ReggeTeitelboim74}. \\
\begin{Def}[Regge--Teitelboim Conditions]
 An initial data set $(M,g,K)$ with an asymptotically Euclidean coordinate chart $\vec{x}$ is said to satisfy the \emph{weak (strong) Regge--Teitelboim conditions} if there exists $\varepsilon>0$ such that, for $\eta=\frac{1}{2}$ ($\eta=1$) and 
  \begin{align}
    g_{ij}^\o &= O_2(r^{-\frac{1}{2}-\eta-\varepsilon})\\
     K_{ij}^\e &= O_1(r^{-\frac{3}{2}-\eta-\varepsilon})\\    
      \mu^\o, J_i^\o &= O_0(r^{-3-\eta-\varepsilon})
\end{align}
as $r=\vert\vec{x}\vert\to\infty$, where the even and odd parts are taken with respect to $\vec{x}$.
 \end{Def}
It was shown by Beig and \'{O} Murchadha in \cite{BOM87} that the strong Regge--Teitelboim conditions indeed suffice to ensure convergence of $\vec{C}_{\text{B\'{O}RT}}$. Consistently, the above Schwarzschild example does not satisfy any Regge--Teitelboim conditions in the asymptotically Euclidean coordinate chart $\vec{y}$ introduced in \eqref{eq:coordchange}, as can be seen by a tedious computation for which we refer the interested reader to \cite{WIP}. 

It is well-known (see \cite{CederbaumNerz2015} and the references cited therein) that the weak Regge--Teitelboim conditions do not suffice to ensure convergence of $\vec{C}_{\text{B\'{O}RT}}$; yet, as we will see at the end of this article, they are very relevant for analyzing $\vec{C}_{\text{B\'{O}RT}}$. 

Moreover, as in the Newtonian Gravity case, neither the strong nor the weak Regge--Teitelboim conditions are invariant under changes between different asymptotically Euclidean coordinate charts because of the same conflict between reflections and translations. But they suffer from even more fundamental issues.

\section{(In-)Existence of Coordinate Systems Satisfying the Regge--Teitelboim Conditions}
We have just seen that the class of coordinate systems satisfying the Regge--Teitelboim conditions is not closed under translations. But, more fundamentally, do all asymptotically Euclidean relativistic initial data sets even possess \emph{any} asymptotically Euclidean coordinate charts in which the (strong) Regge--Teitelboim conditions hold? As we have investigated with Jan Metzger in~\cite{WIP}, this turns out not to be the case; indeed, we will soon give explicit counter-examples.

In order to prove \emph{in}existence of such asymptotically Euclidean coordinate charts on a given relativistic initial data set, we utilize Bartnik's harmonic asymptotically Euclidean coordinate charts, see \Cref{sec:harmonic}, and methods from  \cite{Bartnik86,LeeParker87} as well as a bootstrapping argument to show the following result. We refer the interested reader to our joint work with Jan Metzger~\cite{WIP} for more details and the proofs of the following theorems.

\begin{Thm}\label{thm:harmonicRT}
Let  $(M,g,K)$ be an asymptotically Euclidean relativistic initial data set and assume it satisfies the weak (strong) Regge--Teitelboim conditions with respect to an asymptotically Euclidean coordinate chart $\vec{x}$. Then there exists a smooth harmonic asymptotically Euclidean coordinate chart $\vec{y}$ such that 
    $\vec{x}-\vec{y}= O_3(|\vec{x}|^{\frac{1}{2}-\overline{\varepsilon}})$ as $\vert\vec{x}\vert\to\infty$ and 
    \begin{align}\label{eq:redDerRT1}
\overline{g}_{ij}^{\,\mathrm{odd}} &= O_{1}(\vert\vec{y}\vert^{-\frac{1}{2}-\eta-\overline{\varepsilon}})\\\label{eq:redDerRT2}
\overline{K}_{ij}^{\,\mathrm{even}}  &=O_{0}(\vert\vec{y}\vert^{-\frac{3}{2}-\eta-\overline{\varepsilon}})
\end{align}
as $\vert\vec{y}\vert\to\infty$ for some $\overline{\varepsilon}>0$, where $\eta=\frac{1}{2}$ (respectively $\eta=1$), and where the components $\overline{g}_{ij}$ and $\overline{K}_{ij}$ of $g$ and $K$ as well as their odd and even parts are computed with respect to~$\vec{y}$.
\end{Thm}
In other words, the Regge--Teitelboim conditions are inherited by harmonic asymptotically Euclidean coordinate charts up to a potential loss of derivatives. As a corollary of this analysis, the reduced derivative weak (respectively strong) Regge--Teitelboim conditions \eqref{eq:redDerRT1}, \eqref{eq:redDerRT2} are satisfied for one set of harmonic asymptotically Euclidean coordinate charts if and only if they are satisfied for all such charts. 

We also get the following ``converses'', which readily follow from a more careful analysis of decay rates.

\begin{Thm}\label{thm:notRTinHamonic}
Let  $(M,g,K)$ be an asymptotically Euclidean relativistic initial data set with respect to an asymptotically Euclidean coordinate chart $\vec{x}$, but assume that
\begin{align}
K_{ij}^\e \neq O_0(|\vec{x}|^{-2-\varepsilon})
\end{align}
 as $\vert\vec{x}\vert\to\infty$ for some decay parameter $\varepsilon>0$.
 
Then the harmonic coordinate chart constructed in  \Cref{thm:harmonicRT} is asymptotically Euclidean but cannot satisfy the weak Regge--Teitelboim conditions. More precisely, we get
\begin{align}
\overline{K}_{ij}^{\,\mathrm{even}} \neq  O_0(|\vec{y}|^{-2-\varepsilon})
\end{align}
as $\vert\vec{y}\vert\to\infty$. If, in addition, $g$ satisfies additional decay assumptions such as for example
 \begin{align}
 g^\o_{ij}= O_2(|\vec{x}|^{-\sfrac{3}{2}-\varepsilon}),
 \end{align}
while 
\begin{align}
K_{ij}^\e \neq O_0(|\vec{x}|^{-\frac{5}{2}-\varepsilon})
\end{align}
as $\vert\vec{x}\vert\to\infty$ then the harmonic coordinate chart constructed in \Cref{thm:harmonicRT} is asymptotically Euclidean but cannot satisfy the strong Regge--Teitelboim conditions. More precisely, we get
\begin{align}
\overline{K}_{ij}^{\,\mathrm{even}} \neq  O_0(|\vec{y}|^{-\frac{5}{2}-\varepsilon})
\end{align}
as $\vert\vec{y}\vert\to\infty$.
\end{Thm}

In a nutshell, we have seen that ruling out the existence of asymptotically Euclidean coordinate charts in which a given relativistic initial data set satisfies the strong or the weak Regge--Teitelboim conditions can be reduced to asking (almost) the same question only about harmonic asymptotically Euclidean coordinate charts. 

This allows us to give a number of explicit examples of relativistic initial data sets not allowing for any asymptotically Euclidean coordinate charts satisfying the strong (respectively weak) Regge--Teitelboim conditions.
 
\subsection{Graphical Counter-Examples to Existence of Regge--Teitelboim Coordinates}\label{sec:examples}
\begin{figure}[h]
\begin{center}
\includegraphics[scale=1.1]{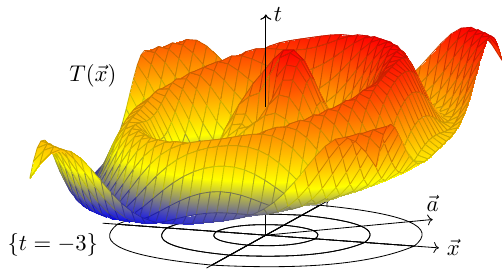}
\caption{Graphical example in Schwarzschild spacetime for $\{t=T_{1}\}$, logarithmic plot.\label{fig:Schwarz}}
\end{center}
\end{figure}

All examples discussed in this section originate from our joint work with Jan Metzger~\cite{WIP}. Following ideas by Cederbaum and Nerz~\cite{CederbaumNerz2015}, we focus on  relativistic initial data sets in the Schwarzschild spacetime of mass $m\in\R$ in the Cartesian coordinates $\vec{x}$ associated with the Schwarzschild coordinates. These can and will be described as graphs over the canonical  relativistic initial data set $\{t=0\}$ of suitable graph functions $T\colon \R^{3}\setminus C\to \R $ for a suitable compact set $C\subset\R^{3}$. Writing the Schwarzschild spacetime as 
\begin{align}
N(r)&=\sqrt{1-\frac{2m}{r}},\\
h&=\frac{1}{N^{2}}dr^{2}+r^{2}d\Omega^{2},
\end{align}
on $\R\times(r_m,\infty)\times\mathbb{S}^2$ with $d\Omega^2$ denoting the canonical metric on the sphere $\mathbb{S}^2$, one finds
\begin{align}
    g^T_{ij}&=h_{ij} -N^2\partial_iT\partial_jT\\
    K^T_{ij}&=\frac{\partial_iT\partial_jN+\partial_jT\partial_iN+N \mathrm{Hess}_h(T)_{ij}-N^2\partial_iT\partial_jT\,dN( \mathrm{grad}_h(T))}{\sqrt{1-N^2|dT|^2_h}}
\end{align}
on the graph $M_T=\{t=T(\vec{x}): \vec{x}\in \R^{3}\setminus C\}$, see \Cref{fig:Schwarz}. \newpage

Choosing 
\begin{align}\label{eq:T1}
T_{1}(\vec{x}) =\sin(\ln r)+\frac{\vec{u}\cdot \vec{x}}{r}
\end{align}
as in \cite{CederbaumNerz2015} for non-trivial $\vec{u}\in\R^{3}$, one obtains a relativistic initial data set $(M_{T_{1}},g^{T_{1}},K^{T_{1}})$ which satisfies neither the weak nor the strong Regge--Teitelboim conditions with respect to~$\vec{x}$; in fact, $\vec{C}_{\text{B\'{O}RT}}$ diverges like $\sin(\ln r)\vec{u}$ in this example, see~\cite{CederbaumNerz2015,WIP}. It is worth noting that the metric $g^{T_{1}}$ in fact \emph{does} satisfy the weak (but not the strong) Regge--Teitelboim conditions with respect to $\vec{x}$ (see \cite{CederbaumNerz2015}); they fail to hold only for~$K^{T_{1}}$.

Suitably exploiting Theorems \ref{thm:harmonicRT} and \ref{thm:notRTinHamonic} and the decay of $T_{1}$, $N$, and $h$, one can assert that $(M_{T_{1}},g^{T_{1}},K^{T_{1}})$ does not satisfy the weak nor the strong Regge--Teitelboim conditions in \emph{any} asymptotically Euclidean coordinate chart. 

Similarly, choosing 
\begin{align}
T_{2}(\vec{x})=\frac{\sin(\ln r)}{r^\beta}
\end{align}
for $0<\beta<\frac{1}{2}$, one finds that $(M_{T_{2}},g^{T_{2}},K^{T_{2}})$ \emph{does} satisfy the weak Regge--Teitelboim conditions (for $\varepsilon<\frac{1}{2}$) but does not possess \emph{any} asymptotically Euclidean coordinate chart in which the strong Regge--Teitelboim conditions hold. Again, the problematic (non-)decay occurs in $K^{T_{2}}$.
 
\subsection{Why the Weak Regge--Teitelboim Conditions are Relevant for the Center of Mass}
Finally, we still owe the reader a justification of why the weak Regge--Teitelboim conditions are relevant for the study of the center of mass $\vec{C}_{\text{B\'{O}RT}}$: Indeed, Huisken and Yau in \cite{HuiskenYau96} developed an alternative definition of center of mass, called $\vec{C}_{\text{CMC}}$, via asymptotic Constant Mean Curvature (CMC) foliations. In a series of works culminating in a paper by Nerz \cite{Nerz2015}, it was shown that, for asymptotically Euclidean relativistic initial data sets satisfying the weak Regge--Teitelboim conditions, one has 
\begin{align}
\vec{C}_{\text{B\'{O}RT}}=\vec{C}_{\text{CMC}}
\end{align}
in the sense that either both centers diverge or both converge to the same limit. 

\begin{figure}[h]
\begin{center}
\includegraphics[scale=0.7]{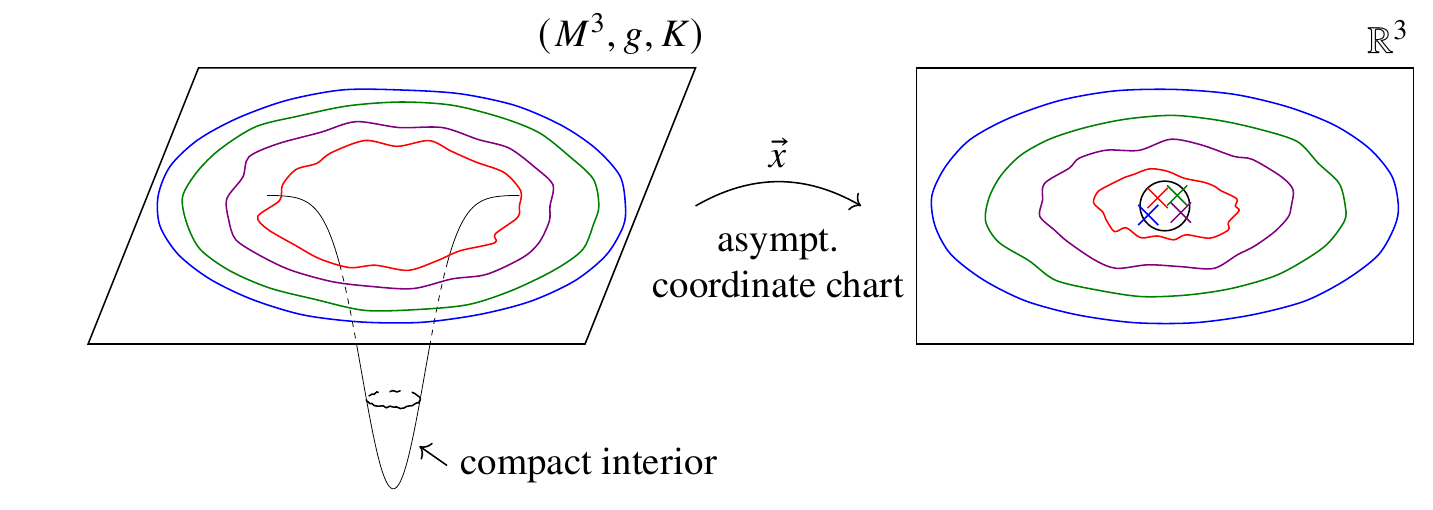}
\end{center}
\caption{The leaves of a foliation near infinity and their images in $\R^{3}$ under the asymptotic coordinates~$\vec{x}$. The crossed positions indicate the Euclidean coordinate centers $\vec{c}\,(\Sigma_{\sigma})$ of the surfaces $\Sigma_{\sigma}$ of the same color (representing the same parameter $\sigma$).}\label{fig:coord-CoM}
\end{figure}

Roughly, Huisken and Yau in \cite{HuiskenYau96} and Nerz in \cite{Nerz2015} prove existence and uniqueness of a foliation (that is, a smoothly parametrized partition into smooth $2$-spheres parametrized by $\sigma\in(0,\sigma_{0})$) of the asymptotic end of an asymptotically Euclidean relativistic initial data set, such that the leaves have constant mean (i.e., average extrinsic) curvature $H(\Sigma_\sigma)=\sigma$. The leaves (i.e., the $2$-spheres) $\Sigma_{\sigma}$ of this foliation are indicated as colored curves in \Cref{fig:coord-CoM}. Pushing forward the leaves via the asymptotic coordinates~$\vec{x}$, $\vec{x}(\Sigma_{\sigma})$, gives rise to a foliation of a neighborhood of infinity in $\R^{3}$ and one computes the average position of a point on $\vec{x}(\Sigma_{\sigma})$ in $\R^{3}$ as
\begin{align}
\vec{c}(\Sigma_{\sigma})=\frac{1}{\vert\vec{x}(\Sigma_{\sigma})\vert}\iint_{\vec{x}(\Sigma_{\sigma})}\vec{x}\,dA_{\delta},
\end{align}
where $\vert\vec{x}(\Sigma_{\sigma})\vert$ denotes the surface area of $\vec{x}(\Sigma_{\sigma})$ in $\R^{3}$ with respect to the Euclidean metric $\delta$, see \Cref{fig:coord-CoM}. The center $\vec{C}_{\text{CMC}}$ then arises as the limit 
\begin{align}
\vec{C}_{\text{CMC}}=\lim_{\sigma\to0}\vec{c}(\Sigma_{\sigma})
\end{align}
outward along this foliation, provided this limit exists. We refer the interested reader to \cite{CederbaumNerz2015} for more information on this construction and its dependence on the choice of asymptotic coordinates. 

\subsubsection{Spacetime Equivariance}
It was observed by Cederbaum and Sakovich in \cite{CederbaumSakovich2021} that the divergence issue of both notions of center of mass for $(M_{T_{1}},g^{T_{1}},K^{T_{1}})$ --- i.e., the one defined via a Hamiltonian systems approached by Beig  and \'{O} Murchadha in \cite{BOM87} and the one defined via foliations --- is rooted in the lack of dependence on $K$ in both approaches. Generalizing the Constant Mean Curvature foliation approach, they construct asymptotic "Spacetime Constant Mean Curvature (STCMC)" foliations in asymptotically Euclidean relativistic initial data sets, see below. These allow for the definition of a generally covariant center of mass $\vec{C}_{\text{STCMC}}$ as well as a correction term $\vec{Z}$ for $\vec{C}_{\text{B\'{O}RT}}$ such that 
\begin{align}
\vec{C}_{\text{STCMC}}=\vec{C}_{\text{B\'{O}RT}}+\vec{Z}
\end{align}
holds under the weak Regge--Teitelboim conditions in the sense that either both sides of the equation diverge or both converge to the same limit. The definition of $\vec{C}_{\text{STCMC}}$ mimicks the definition of $\vec{C}_{\text{CMC}}$, see \Cref{fig:coord-CoM}, based on the Spacetime Constant Mean Curvature instead of on the (spatial) Constant Mean Curvature foliation. As expected from the spacetime symmetry, one finds $\vec{C}_{\text{STCMC}}=\vec{0}$ in the graphical example for the graph function $T_{1}$ in \eqref{eq:T1}.

It is also proved in \cite{CederbaumSakovich2021} that $\vec{C}_{\text{STCMC}}$ evolves in time such that
\begin{align}
\frac{d}{dt}\vec{C}_{\text{STCMC}}=\frac{\vec{P}_{\text{ADM}}}{E_{\text{ADM}}}
\end{align}
under the Einstein Evolution Equations, just as a freely falling point particle in Special Relativity. This applies even when $\vec{C}_{\text{STCMC}}$ does not converge.

Before we end this section, let us briefly address what it means that a surface $\Sigma$ has "constant spacetime mean curvature": If one considers a $2$-surface $\Sigma$ not only as sitting inside the  relativistic initial data set $(M,g,K)$ but also as sitting inside the spacetime generated from this relativistic initial data set via the Einstein Evolution Equations then it can be viewed as a co-dimension $2$ surface in this spacetime. As such, it has co-dimension $2$ extrinsic curvature, taking the form of a normal vector valued symmetric $(0,2)$-tensor field. The trace (or average) of this normal vector valued symmetric $(0,2)$-tensor field is called the \emph{spacetime mean curvature vector (field) $\overrightarrow{\mathcal{H}}$} of $\Sigma$. The Lorentzian length of $\overrightarrow{\mathcal{H}}$, $\mathcal{H}$, is called the \emph{spacetime mean curvature of $\Sigma$}. Then, a \emph{spacetime constant mean curvature surface} is a surface with $\mathcal{H}=\text{const}$. It turns out that one can compute $\mathcal{H}$ from the initial data alone, without any reference to the ambient spacetime, and one finds
\begin{align}
\mathcal{H}=\sqrt{H^2- (\operatorname{tr}_\Sigma K)^2},
\end{align}
where $\operatorname{tr}_{\Sigma}K$ is the (partial) trace of $K$ over $\Sigma$ and $H$ denotes the (spatial) mean curvature of $\Sigma$ within the initial data set already considered in the Constant Mean Curvature foliation suggested by Huisken and Yau. On the other hand, taking a more physical perspective, one finds that
\begin{align}
\mathcal{H}^2=\theta_+\theta_-,
\end{align}
where $\theta_{\pm}$ denote the null expansions of $\Sigma$ in the ambient spacetime.

\section{Lessons Learned and Current Research}
We have seen that coordinates are messy in the following sense: In Newtonian Gravity, when ahistorically considering general asymptotically Euclidean coordinate charts on $\R^{3}\setminus C$ outside some compact set $C$, convergence of the center of mass depends not only on suitable decay of the matter density but also on the choice of coordinate system. Accordingly, in General Relativity, where we generally do not have preferred systems of coordinates, one cannot hope to have convergence of any notion of center of mass in \emph{all} asymptotically Euclidean coordinate charts. It was suggested to remedy such divergence issues by resorting to asymptotic parity conditions, however, as we showed, not all asymptotically Euclidean relativistic initial data sets have asymptotic parity.

It hence remains an open question whether, instead of asking for asymptotic parity, one can find a condition on asymptotic coordinate charts which is geometric (i.e., coordinate independent) just as Bartnik's harmonic coordinates, compatible with translations (and reflections), and implies convergence of $\vec{C}_{\text{STCMC}}$, and of course such that every asymptotically Euclidean relativistic initial data set carries such a coordinate system. Coordinate charts satisfying such a condition could then legitimately be considered a natural analog of Cartesian coordinates in General Relativity. This question is currently studied by the authors and our coauthor Jan Metzger.

\paragraph{Acknowledgements.} This work was supported by the focus program on Geometry at Infinity (Deutsche Forschungsgemeinschaft,  SPP 2026). MG also acknowledges support by the Deutsche Forschungsgemeinschaft (DFG, German Research Foundation) under Germany's Excellence Strategy -- EXC 2121 ``Quantum Universe'' -- 390833306. The authors would like to thank Axel Fehrenbach, Felix Salfelder, Oliver Schoen, Olivia Vi\v{c}\'{a}nek Mart\'{i}nez, and Giorgos Vretinaris for help with the graphics.

\bibliographystyle{amsplain}
\bibliography{references-Cederbaum}

\providecommand{\bysame}{\leavevmode\hbox to3em{\hrulefill}\thinspace}
\providecommand{\MR}{\relax\ifhmode\unskip\space\fi MR }
\providecommand{\MRhref}[2]{%
  \href{http://www.ams.org/mathscinet-getitem?mr=#1}{#2}
}
\providecommand{\href}[2]{#2}
\begin{thebibliography}{10}

\bibitem{ADM1962}
Richard Arnowitt, Stanley Deser, and Charles~W. Misner, \emph{The dynamics of
  general relativity}, Gravitation: {A}n introduction to current research,
  Wiley, New York, 1962, pp.~227--265.

\bibitem{Bartnik86}
Robert Bartnik, \emph{The mass of an asymptotically flat manifold}, Comm. Pure
  Appl. Math. \textbf{39} (1986), no.~5, 661--693.

\bibitem{BOM87}
Robert Beig and Niall \'{O}~Murchadha, \emph{The {P}oincar\'{e} group as the
  symmetry group of canonical general relativity}, Ann. Physics \textbf{174}
  (1987), no.~2, 463--498.

\bibitem{WIP}
Carla Cederbaum, Melanie Graf, and Jan Metzger, \emph{Initial data sets that do
  not satisfy the {R}egge--{T}eitelboim conditions}, 2023, Work in Progress.

\bibitem{CederbaumNerz2015}
Carla Cederbaum and Christopher Nerz, \emph{Explicit {R}iemannian manifolds
  with unexpectedly behaving center of mass}, Ann. Henri Poincar\'{e}
  \textbf{16} (2015), no.~7, 1609--1631.

\bibitem{CederbaumSakovich2021}
Carla Cederbaum and Anna Sakovich, \emph{On center of mass and foliations by
  constant spacetime mean curvature surfaces for isolated systems in general
  relativity}, Calc. Var. Partial Differential Equations \textbf{60} (2021),
  no.~6, Paper No. 214, 57.

\bibitem{DenSol1983}
Viktor~I. Denissov and Vladimir~O. Solovyev, \emph{The energy determined in
  general relativity on the basis of the traditional {H}amiltonian approach
  does not have physical meaning}, Theor. Math. Phys. \textbf{56} (1983),
  no.~2, 832--841.

\bibitem{HuiskenYau96}
Gerhard Huisken and Shing-Tung Yau, \emph{Definition of center of mass for
  isolated physical systems and unique foliations by stable spheres with
  constant mean curvature}, Invent. Math. \textbf{124} (1996), no.~1-3,
  281--311.

\bibitem{LeeParker87}
John~M. Lee and Thomas~H. Parker, \emph{The {Y}amabe problem}, Bull. Amer.
  Math. Soc. (N.S.) \textbf{17} (1987), no.~1, 37--91.

\bibitem{Nerz2015}
Christopher Nerz, \emph{Foliations by stable spheres with constant mean
  curvature for isolated systems without asymptotic symmetry}, Calc. Var.
  Partial Differential Equations \textbf{54} (2015), no.~2, 1911--1946.

\bibitem{ReggeTeitelboim74}
Tullio Regge and Claudio Teitelboim, \emph{Role of surface integrals in the
  {H}amiltonian formulation of general relativity}, Ann. Physics \textbf{88}
  (1974), 286--318.

\bibitem{SchoenYau}
Richard Schoen and Shing-Tung Yau, \emph{On the proof of the positive mass
  conjecture in general relativity}, Comm. Math. Phys. \textbf{65} (1979),
  no.~1, 45--76.

\bibitem{Witten}
Edward Witten, \emph{A new proof of the positive energy theorem}, Comm. Math.
  Phys. \textbf{80} (1981), no.~3, 381--402.

\end{thebibliography}
\end{document}